\documentclass[aps,prb,twocolumn,superscriptaddress]{revtex4-1}
\usepackage{natbib}
\usepackage{graphicx,amssymb,amsmath,ifthen,braket,xcolor}
\usepackage{bm}

\newcommand{\angstrom}{\mbox{\normalfont\AA}}

\begin{document}


\title{Quantized zero bias conductance plateau in semiconductor-superconductor heterostructures without non-Abelian Majorana zero modes}


\author{Christopher Moore}
\affiliation{Department of Physics and Astronomy, Clemson University, Clemson, SC 29634, USA}

\author{Chuanchang Zeng}
\affiliation{Department of Physics and Astronomy, Clemson University, Clemson, SC 29634, USA}

\author{Tudor D. Stanescu}
\affiliation{Department of Physics and Astronomy, West Virginia University, Morgantown, WV 26506, USA}

\author{Sumanta Tewari}
\affiliation{Department of Physics and Astronomy, Clemson University, Clemson, SC 29634, USA}



\date{\today}

\begin{abstract}
We show that partially separated Andreev bound states (ps-ABSs), comprised of pairs of overlapping Majorana bound states (MBSs) emerging in quantum dot-semiconductor-superconductor heterostructures, produce robust zero bias conductance plateaus in end-of-wire charge tunneling experiments. These plateaus remain quantized at $2e^2/h$ over large ranges of experimental control parameters.
In light of recent experiments reporting the observation of robust $2e^2/h$-quantized conductance plateaus in local charge tunneling experiments, we 
perform extensive numerical calculations to explicitly show that such quantized conductance plateaus, which are obtained by varying control parameters such as the tunnel barrier height, the super gate potential, and the applied magnetic field, can arise as a result of the existence of ps-ABSs. Because ps-ABSs can form rather generically in the topologically trivial regime, even in the absence of disorder, our results suggest that the observation of a robust quantized conductance plateau does not represent sufficient evidence to demonstrate the existence of non-Abelian topologically-protected Majorana zero modes localized at the opposite ends of a wire.
\end{abstract}


\maketitle

%
\section{Introduction}\label{sec:intro}
%
Semiconductor nanowires with proximity induced superconductivity and strong Rashba spin-orbit coupling, which are predicted\cite{sau2010generic,tewari2010theorem,alicea2010majorana,sau2010non,lutchyn2010majorana,oreg2010helical,stanescu2011majorana} theoretically to support mid-gap  non-Abelian Majorana bound states, also called Majorana zero modes (MZMs),\cite{read2000paired,kitaev2001unpaired,nayak2008non,beenakker2013search,elliott2015colloquium} have become the leading candidate for the realization of a topological quantum qubit, mostly due to the tremendous experimental progress realized in past few years.\cite{mourik2012signatures,deng2012anomalous,das2012zero,rokhinson2012fractional,churchill2013superconductor,finck2013anomalous,albrecht2016exponential,deng2016majorana,zhang2017ballistic,chen2017experimental,nichele2017scaling,zhang2018quantized} The most recent relevant development has been the observation of the $2e^2/h$ zero-bias quantized\cite{sengupta2001midgap,akhmerov2009electrically,law2009majorana,flensberg2010tunneling} conductance plateau.\cite{zhang2018quantized}. While previous theoretical work on proximitized semiconducting nanowires has shown the formation of robust ZBCPs before the system undergoes a topological quantum phase transition (TQPT), due to disorder,\cite{bagrets2012class,liu2012zero,degottardi2013majoranaprl,degottardi2013majoranaprb,rainis2013towards,adagideli2014effects} non uniform parameters,\cite{kells2012near,chevallier2012mutation,roy2013topologically,san2013multiple,ojanen2013topological,stanescu2014nonlocality,cayao2015sns,klinovaja2015fermionic,san2016majorana,fleckenstein2017decaying} weak antilocalization,\cite{pikulin2012zero} and coupling to a quantum dot,\cite{prada2012transport,liu2017andreev} these peaks do not typically result in a $2e^2/h$--quantized conductance plateau that is robust against variations of various control parameters, which is taken as a key signature of non-Abelian topologically-protected MZMs localized at the opposite ends of the wire\cite{zhang2018quantized}.

In this paper we perform detailed numerical calculations of quantum dot-semiconductor-superconductor (QD-SM-SC) nanowires and show explicitly that, in fact, the emergence of $2e^2/h$--quantized zero-bias conductance peaks (ZBCPs) that are robust over a large range of control parameters is also possible due to the presence of partially separated Andreev bound states (ps-ABSs)\cite{moore2018two-terminal} localized near the end of the hybrid system containing the quantum dot. Following the recent  experiments,\cite{zhang2018quantized} we show that a quantized conductance plateau emerges with increasing the Zeeman field within a large range of chemical potentials. Furthermore, we show that the ZBCP quantization is robust against variations of the tunnel-barrier potential and changes of the quantum dot potential.
We also simulate variations of the super-gate potential, which result in adjustments of the chemical potential throughout the entire wire. Again, we find a robust $2e^2/h$-quantized conductance plateau, similar to the expected signature of non-Abelian MZMs. We also show that a ps-ABS induced ZBCP behaves identically to a MZM generated conductance peak when the Zeeman field is rotated. Finally, we show that both the ps-ABS and the MZM induced conductance peaks scale identically with temperature.

We interpret the results of this study within a framework based on two basic observations: i) MZMs and ps-ABSs can be described theoretically using the same modeling of the hybrid structure. However, in the low Zeeman field regime the ps-ABSs are significantly  more common, because the parameter region corresponding to inhomogeneous systems that support ps-ABSs is much larger than the parameter region associated with nearly homogeneous systems that host MZMs. ii) The goal of this study is {\em not} to identify the nature of the low-energy states responsible for the signatures observed experimentally (much less to demonstrate that these states are ps-ABSs). Given the fundamental uncertainty regarding key parameters of the hybrid systems used in experiments, such as, for example, work function differences and couplings across the SM-SC interface,  any attempt to solve these problems theoretically would be futile. The answer has to come from experiment. Here, we only show that the signature produced by a ps-ABS in a local tunneling measurement is indistinguishable from the corresponding signature of a MZM, \textit{even if we test the robustness of this signature by varying the control parameters}. Based on the results of our numerical analysis, we conclude that the single-terminal charge tunneling measurement (which was, so far, the primary type of probe used  experimentally) has exhausted its potential to reveal useful information regarding the nature of the low-energy states in SM-SC hybrid structures. The next stage must involve non-local probes, such as, for example, the two-terminal charge tunneling measurement.\cite{moore2018two-terminal}

The rest of this paper is structured as follows: In Sec \ref{sec:SMSC} we describe the numerical model used throughout the paper. In Sec. \ref{sec:results} we show explicitly the formation of $2e^2/h$-quantized conductance plateaus as different experimentally--relevant parameters are varied. We end with a brief conclusion in Sec. \ref{sec:conclusion}.

%
\section{SM-SC Heterostructure coupled to a quantum dot}\label{sec:SMSC}
%
\begin{figure}
	\begin{center}
		\includegraphics[width=0.48\textwidth]{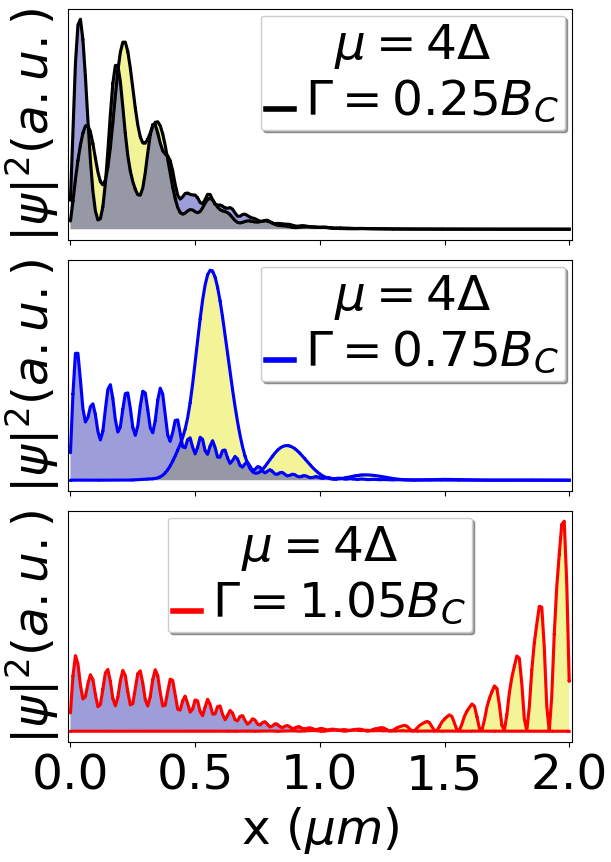}\llap{\parbox[b]{180mm}{\large\textbf{(a)}\\\rule{0ex}{115mm}}}\llap{\parbox[b]{180mm}{\large\textbf{(b)}\\\rule{0ex}{80mm}}}\llap{\parbox[b]{180mm}{\large\textbf{(c)}\\\rule{0ex}{45mm}}}
	\end{center}
	\caption{(Color online) Profiles of Majorana bound state wave functions associated with low energy spectrums in Fig \ref{fig:potential}(e). (a) A standard ABS consisting of a pair of overlapping MBSs, (b) a ps-ABS consisting of two overlapping MBSs whose separation is on the order of the length scale of the Majorana decay length $\zeta$, and  (c) a pair of MZMs localized at each end of the wire.}
	\label{fig:wfPlats}
\end{figure}
\begin{figure*}
	\begin{center}
		\includegraphics[width=0.42\textwidth]{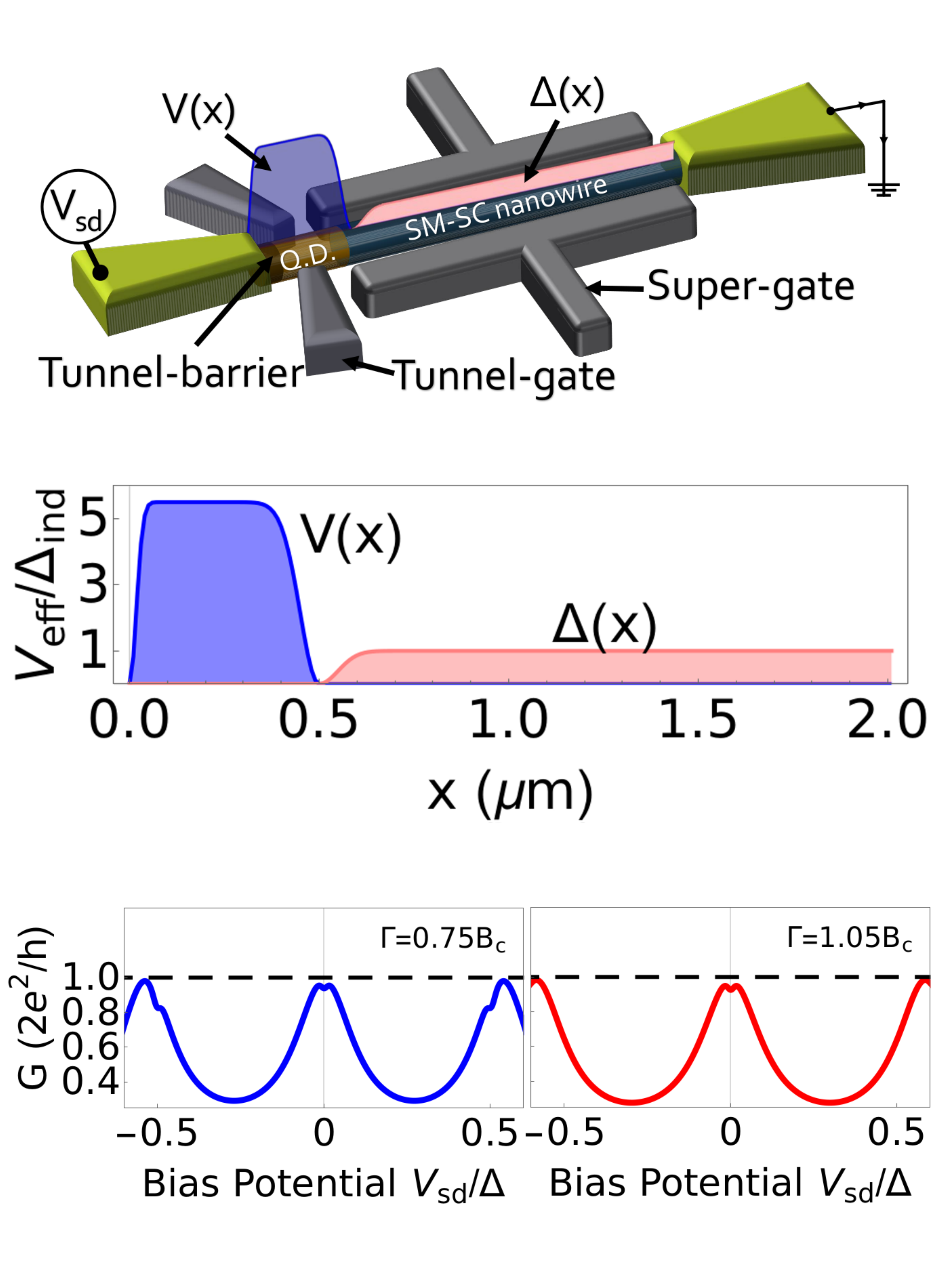}
		\llap{\parbox[b]{125mm}{\large\textbf{(a)}\\\rule{0ex}{98mm}}}
		\llap{\parbox[b]{125mm}{\large\textbf{(b)}\\\rule{0ex}{50mm}}}
		\llap{\parbox[b]{128mm}{\large\textbf{(c)}\\\rule{0ex}{25mm}}}\llap{\parbox[b]{64mm}{\large\textbf{(d)}\\\rule{0ex}{25mm}}}		
		\includegraphics[width=0.55\textwidth]{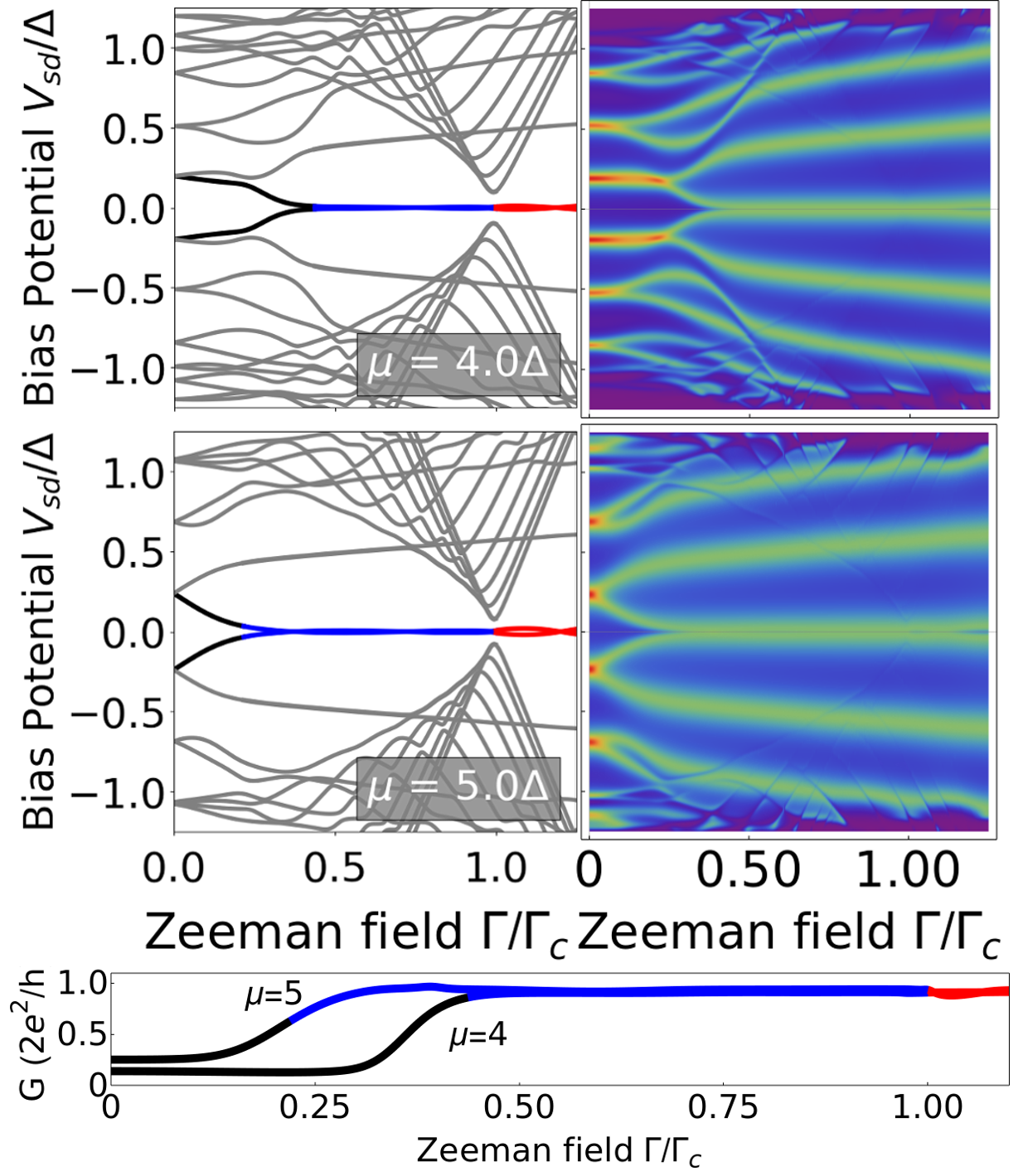}
		\llap{\parbox[b]{155mm}{\Large\textbf{(e)}\\\rule{0ex}{106mm}}}\llap{\parbox[b]{77mm}{{\Large\textbf{\textcolor{white}{(g)}}}\\\rule{0ex}{106mm}}}\llap{\parbox[b]{155mm}{\Large\textbf{(f)}\\\rule{0ex}{65mm}}}\llap{\parbox[b]{77mm}{{\Large\textbf{\textcolor{white}{(h)}}}\\\rule{0ex}{65mm}}}\llap{\parbox[b]{25mm}{\Large\textbf{(i)}\\\rule{0ex}{10mm}}}
		
	\end{center}
	\caption{(a) Proximitized semiconductor NS nanowire junction in which a portion of the semiconductor wire (SM) is not covered by the superconductor (SC), represented by a quantum dot (QD). (b) Potential profile $V(x)$ may form within the QD due to tunnel coupling between the normal lead and the SM wire, differences in the work functions in the QD and the SM-SC regions, as well as application of a tunnel gate potential. Induced pairing $\Delta_{ind}=0.25$meV within the proximitized region $\Delta(x)$. (c)-(d) Vertical line cuts from (g) showing a ZBCP quantized to $2e^2/h$ due to the presence of a ps-ABS (blue) and an MZM (red).  (e)-(f) Low-energy spectra as a function of Zeeman field for a nanowire associated with the potential profile pictured in (b). The bulk critical field $\Gamma>\Gamma_c=\sqrt{\Delta^2+\mu^2}$ is marked by the red zero energy mode, while the blue zero mode marks the region supporting trivial ps-ABSs. (g)-(h) Differential conductance spectra as a function of Zeeman field corresponding to energy spectra in (e) and (f) respectively.  (i) Zero bias line cuts from (g) and (h) showing $2e^2/h$-quantized conductance plateaus forming before the TQPT due to the presence of a ps-ABS.}
	\label{fig:potential}
\end{figure*}

We consider a semiconducting nanowire proximity coupled to a superconductor with strong spin orbit coupling in the presence of an applied magnetic field, in which a portion of the SM wire is not in contact with the SC as shown in Fig. \ref{fig:potential}(a). The portion of the SM wire which is not in contact with the SC may be thought of as a quantum dot coupled to the end of a SM-SC heterostructure.\cite{prada2012transport,liu2017andreev,moore2018two-terminal} Such quantum dots often form at the ends of the wire when gating nanowires.

The Bogoliubov-de Gennes (BdG) Hamiltonian for a one dimensional semiconductor-superconductor heterostructure can be written as
\begin{align}
	\begin{split}
		&\tilde{H}_{NW}=\left(-\frac{1}{2}\partial_{\tilde{x}}^2-i\partial_{\tilde{x}}\sigma_y-\tilde{\mu}+V\left(\tilde{x}\right)\right)\tau_z+\Gamma_{\tilde{x}}\sigma_x+\Delta\left(\tilde{x}\right)\tau_x
		\\
		&\tilde{H}_{Lead}=\left(-\frac{1}{2}\partial_{\tilde{x}}^2-i\partial_{\tilde{x}}\sigma_y-\tilde{\mu}+V\left(\tilde{x}\right)\right)\tau_z+\Gamma_{\tilde{x}}\sigma_x
	\end{split}
	\label{eq:Hamiltonian}
\end{align}
with $\tilde{x}=m^\ast\alpha x$ and $\tilde{E}=(E/m^\ast \alpha^2)$. Here $\sigma_i$ and $\tau_j$ are the Pauli matrices operating in spin and particle-hole space respectively, $\Gamma$ is the Zeeman field and $\mu$ is the chemical potential. The parameters for this tight binding model correspond to an effective mass $m^\star \approx 0.03m_0$ with an induced gap of $\Delta=0.25$ meV, and a Rashba coefficient of $\alpha = 400$ meV$\cdot\angstrom$. All calculations were done at a temperature $T\approx20$ mK unless otherwise noted. Here $V\left(\tilde{x}\right)$ represents the potential due to the quantum dot as shown in Fig. \ref{fig:potential}(b).

The low-energy spectrum is obtained by numerically diagonalizing the BdG Hamiltonian corresponding to the nanowire. Values for the differential conductance $G$ were found by discretizing the Hamiltonians in Eq. \ref{eq:Hamiltonian} as follows,
\begin{align}
	\begin{split}\label{eq:hamdis}
		\hat{H}_{NW} = \sum_{i}\{&\psi_{i}^\dagger\left[\left(2t-\mu +V\left(i\right)\right)\tau_z +\Gamma\sigma_x + \Delta\left(i\right)\tau_x\right]\psi_i \\
		&+\left[\psi_{i+a}^{\dagger}\left(-t\tau_{z}+i\alpha\sigma_y\tau_x\right)\psi_i + h.c.\right] \}
		\\
		\hat{H}_{Lead} = \sum_{i}&\{\psi_{i}^\dagger\left[\left(2t-\mu\right)\tau_z +\Gamma\sigma_x\right]\psi_i \\
		&+\left[\psi_{i+a}^{\dagger}\left(-t\tau_{z}\right)\psi_i + h.c.\right] \}
	\end{split}
\end{align}
written in the Nambu basis with $\psi_i = \left(c_{\uparrow i},c_{\downarrow i},c_{\uparrow i}^{\dagger},c_{\downarrow i}^{\dagger}\right)$ in which $i$ represents the lattice site and $t$ is the hopping matrix element. The zero temperature differential conductance
\begin{equation}
	G_0(V) = \frac{e^2}{h} (N - R_\text{ee} + R_\text{he}),
\end{equation}
was found using the S matrix method in KWANT \cite{groth2014kwant}, a Python package for transport calculations for systems with tight binding Hamiltonians. Here $N$ is the number of electron channels in the lead, $R_{ee}$ is the total probability of normal reflection and  $R_{eh}$ is the total probability of Andreev reflection for an electron in the lead. Finite temperature is represented by broadening the zero temperature conductance through a convolution with the derivative of the Fermi-function in the usual manner, $G\left(V,T\right)=-\int d\epsilon G_0\left(\epsilon\right)f_T^\prime\left(\epsilon-V\right)$.
\begin{figure*}
	\begin{center}
		\includegraphics[width=0.49\textwidth]{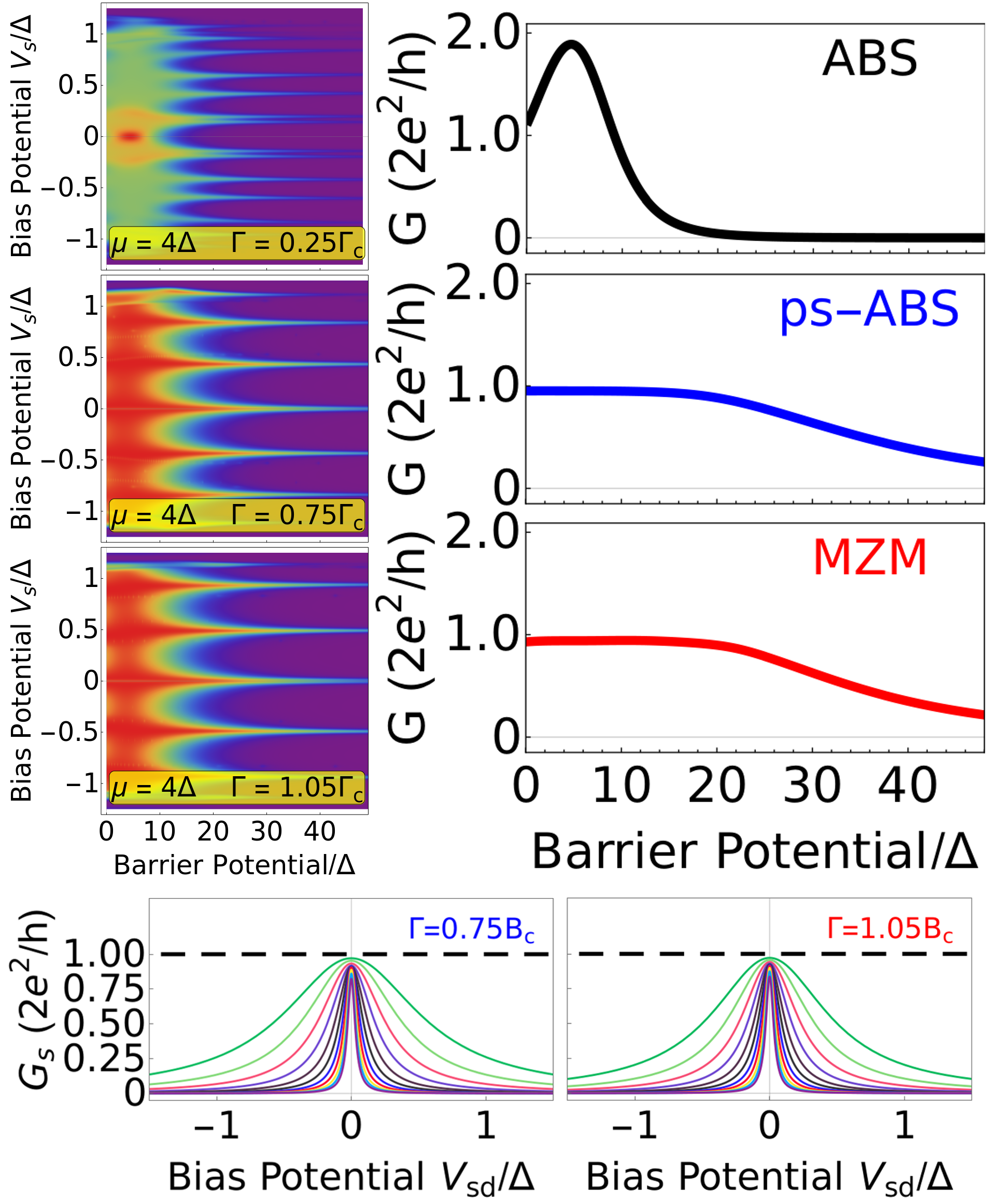}\hspace{3mm}
		\llap{\parbox[b]{158mm}{\large\textbf{\textcolor{white}{(a)}}\\\rule{0ex}{101mm}}}
		\llap{\parbox[b]{85mm}{\large\textbf{(d)}\\\rule{0ex}{85mm}}}\llap{\parbox[b]{160mm}{\large\textbf{\textcolor{white}{(b)}}\\\rule{0ex}{76mm}}}\llap{\parbox[b]{85mm}{\large\textbf{(e)}\\\rule{0ex}{63mm}}}\llap{\parbox[b]{160mm}{\large\textbf{\textcolor{white}{(c)}}\\\rule{0ex}{52mm}}}\llap{\parbox[b]{85mm}{\large\textbf{(f)}\\\rule{0ex}{41mm}}}
		\llap{\parbox[b]{153mm}{\large\textbf{(g)}\\\rule{0ex}{22mm}}}\llap{\parbox[b]{78mm}{\large\textbf{(h)}\\\rule{0ex}{22mm}}}
		\includegraphics[width=0.47\textwidth]{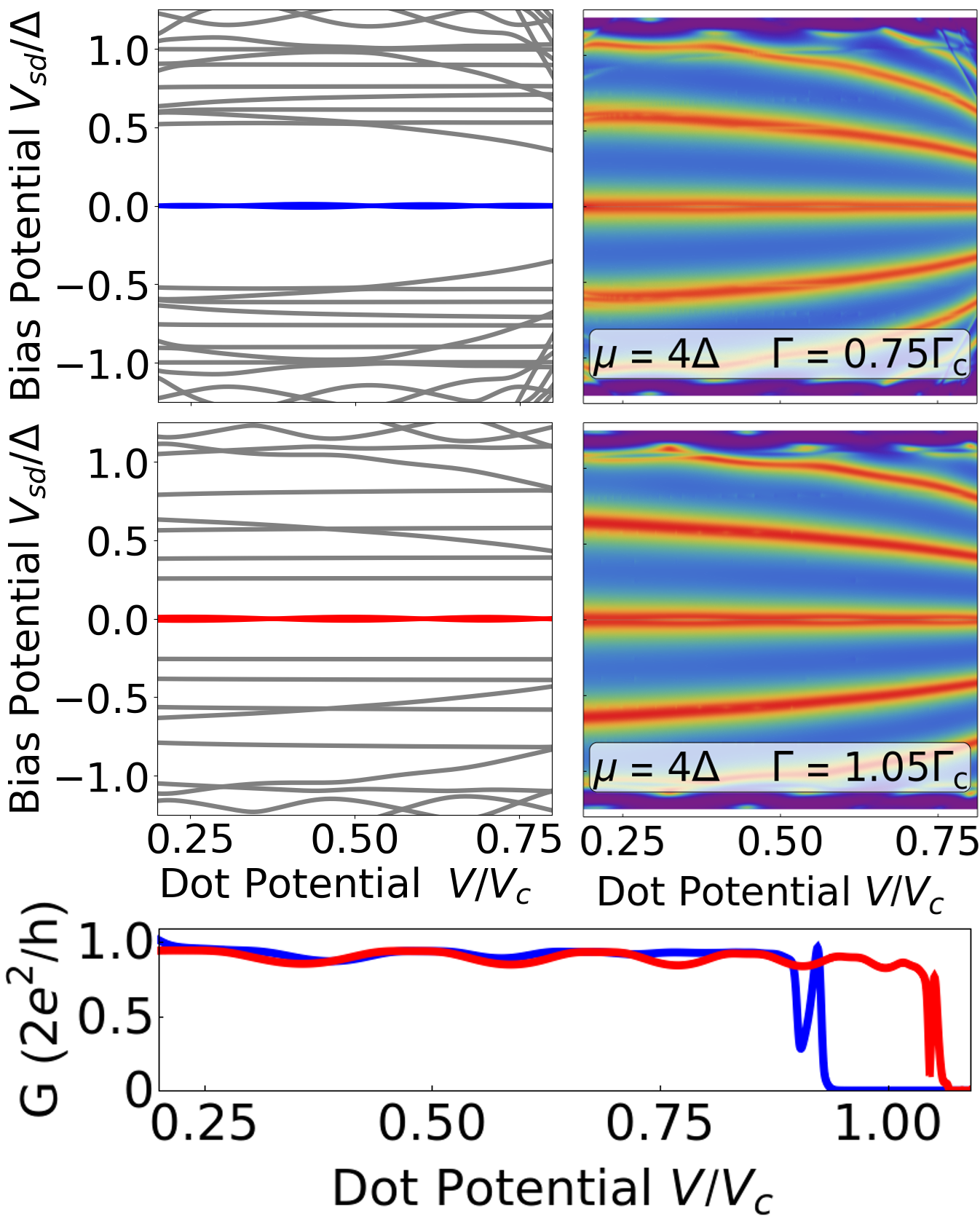}\llap{\parbox[b]{135mm}{\large\textbf{(i)}\\\rule{0ex}{98mm}}}\llap{\parbox[b]{62mm}{\large\textbf{\textcolor{white}{(k)}}\\\rule{0ex}{98mm}}}\llap{\parbox[b]{135mm}{\large\textbf{(j)}\\\rule{0ex}{63mm}}}\llap{\parbox[b]{63mm}{\large\textbf{\textcolor{white}{(l)}}\\\rule{0ex}{63mm}}}\llap{\parbox[b]{133mm}{\large\textbf{(m)}\\\rule{0ex}{12mm}}}
	\end{center}
	\caption{(Color online) (a)-(c) Differential conductance as a function of barrier height $Z$ and bias potential associated with an ABSs (a), a ps-ABSs (b), and a pair of MZMs (c).(d-f) Zero bias line cuts for (a-c) showing the MZMs and ps-ABSs forming nearly identical profiles which plateau at $2e^2/h$ for a wide range of barrier heights. The ABS may take any value $0$ to $4e^2/h$ and quickly goes to $0$ with increased barrier height. (g-h) Vertical line cuts from (b) and (c) showing ZBCPs quantized at $2e^2/h$ over a large range of barrier potentials $Z$ for both ps-ABS and MZM.  (i-j) Low energy spectra as a function of potential height $V$ associated with potential profile in Fig \ref{fig:potential}(b). (k-l) Plots of differential conductance as a function of potential height $V$ and bias potential for values consistent with energy-spectra shown in (i-j). (m) Horizontal zero bias line cuts taken from (k-l) showing a $2e^2/h$-quantized plateau for both ps-ABSs (blue) and MZMs (red).}
	\label{fig:barPlats}
\end{figure*}
The tunnel barrier arises due to tunnel coupling between the normal lead and the SM wire and is present independent of the existence of the quantum dot. The value of the tunnel barrier height $Z$ is dependent on numerous parameters which may not be available to directly control experimentally.

To characterize the Andreev bound states (ABSs) in the system note that each BdG eigenstate can be represented as a pair of overlapping MBSs. A pair of states, $\chi_A\left(i\right)=\frac{1}{\sqrt{2}}\left[\phi_\varepsilon\left(i\right)+\phi_{-\varepsilon}\left(i\right)\right]$ and $\chi_B\left(i\right)=\frac{i}{\sqrt{2}}\left[\phi_\varepsilon\left(i\right)+\phi_{-\varepsilon}\left(i\right)\right]$, is constructed which are linear combinations of the wave functions $\phi_{\pm}\left(i\right)$ of the low energy states of Eq. \ref{eq:hamdis}. Using this formalism a standard ABS is defined as an ABS in which the constituent MBSs are sitting directly on top of one another Fig. \ref{fig:wfPlats}(a), a ps-ABS as an ABS in which the constituent MBSs are separated on the order of the Majorana decay length $\zeta$  Fig. \ref{fig:wfPlats}(b), and topological MZMs as a state in which the constituent MBSs are separated by the length of the wire Fig. \ref{fig:wfPlats}(c). By plotting the wave function using this formalism it is straight forward to see that if a ps-ABS is present in the quantum dot as in Fig. \ref{fig:wfPlats}(b) a tunnel probe placed on the left hand side of the wire will predominantly couple to a single MBS (purple) making it indistinguishable from an MZM as in Fig \ref{fig:wfPlats}(c). Note that, in a finite wire the bulk gap does not completely close, but only goes through a minimum at the TQPT, thus a ps-ABS can be continuously connected to a pair of topological MZMs. As opposed to an infinite wire in which the ps-ABS and the MZMs are separated by a bulk gap closure signaling a TQPT. Moreover ps-ABSs can not be used in topological quantum computations, because the separation between the MBSs can not be controlled independently.\cite{moore2018two-terminal}


%
\section{Results}\label{sec:results}
%

\begin{figure}
	\begin{center}
		\includegraphics[width=0.48\textwidth]{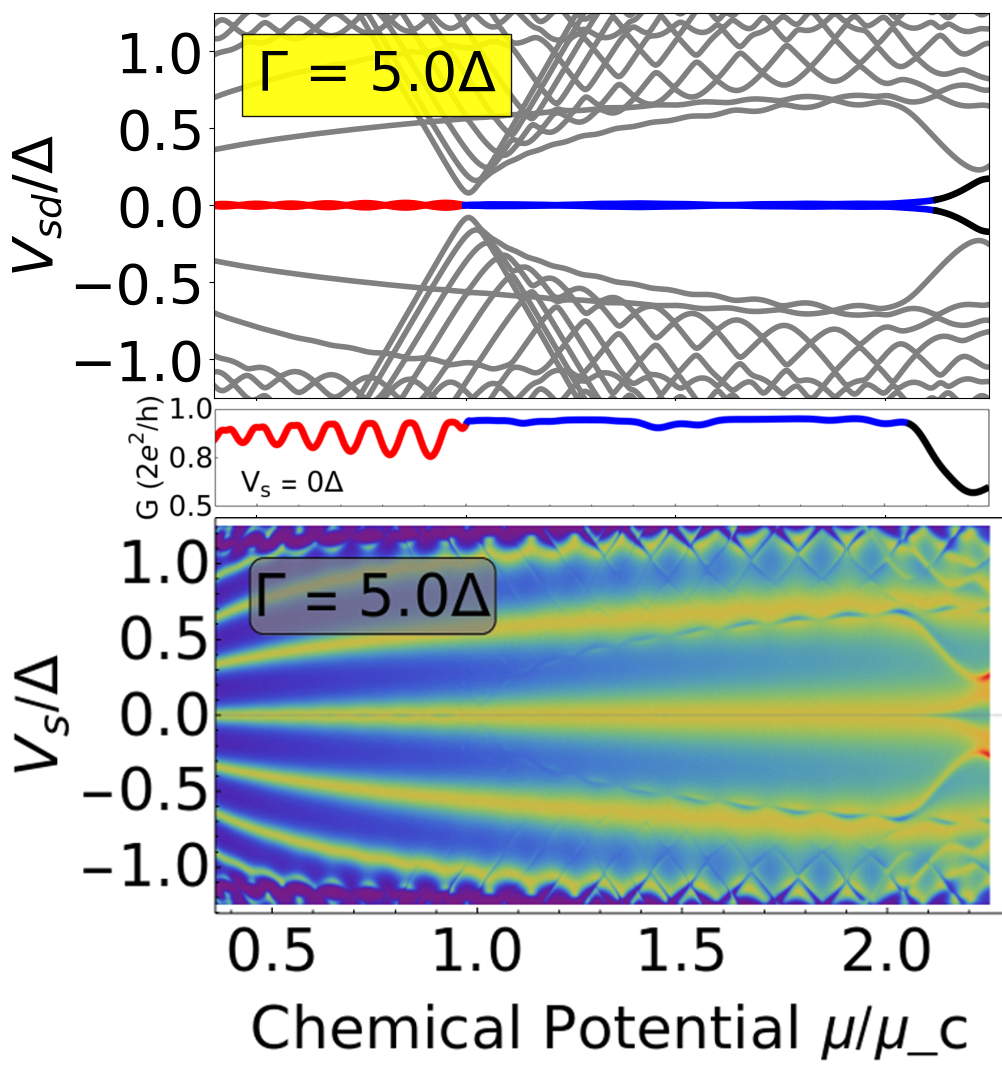}
	\end{center}
	\caption{(Color online)(top)Low-energy spectra as a function of chemical potential. The red line signifies the topological region supporting MZMs, the blue line shows a non-topological region supporting ps-ABSs which stick to zero energy for a wide range of chemical potential. (bottom) Differential conductance spectrum as a function of chemical potential. (middle) ZBCP as a function of chemical potential. Conductance shows dips in the quantized conductance peak due to peak splittings which return to $2e^2/h$, but do not exceed it.}
	\label{fig:Fig4}
\end{figure}
%
\subsection{Quantized Conductance Plateau with Variation of the Zeeman field}\label{sec:zemField}
%



Fig \ref{fig:potential}(e-f) shows the low energy spectra as a function of Zeeman field. A pair of robust zero modes emerge in each plot well before the bulk band gap has a minimum signaling the TQPT. Corresponding plots of differential conductance as a function of Zeeman field (Fig. \ref{fig:potential}(g-h)), show robust conductance plots in the topologically trivial regime which are indistinguishable from those after the bulk band gap minimum. These robust ZBCPs form a $2e^2/h$-quantized conductance plateau in the topologically trivial and topological regime Fig. \ref{fig:potential}(i) similar to those shown in experimental papers.\cite{zhang2018quantized}.

Wave functions $\chi_A$ and $\chi_B$ corresponding to the low energy modes in Fig. \ref{fig:potential}(e) are plotted in Fig. \ref{fig:wfPlats}. The top panel shows a standard ABS, the middle a ps-ABS, and the bottom shows a pair of MZM. The ps-ABS and the MZMs will appear nearly identical to a tunnel probe on the left end of the wire, due to the fact that the tunnel probe will couple to only one of the constituent MBSs. The small overlap between the MBSs in the ps-ABS in Fig. \ref{fig:potential}(b) and the MZMs in Fig. \ref{fig:potential}(c) leads to the $2e^2/h$-quantized conductance plateaus as shown in Fig. \ref{fig:potential}(i).

%
\subsection{Quantized Conductance Plateau with Variation Tunneling barrier height}\label{sec:tunBar}
%

Plots of differential conductance as a function of barrier potential Fig. \ref{fig:barPlats}(a-f) show that while the ZBCP height due to a standard ABS may take any value ($0-4e^2/h$) and quickly drops to zero for increased barrier potential, the differential conductance spectra for ps-ABS and MZMs are nearly identical. The ps-ABS and MZM both form quantized conductance plateaus at $2e^2/h$ in the ZBCP measured with respect to tunneling barrier height $Z$ (Fig. \ref{fig:barPlats}(e-f)), which are nearly indistinguishable from one another, as expected. Fig. \ref{fig:barPlats}(m) shows the ps-ABS (blue) and MZM (red) have nearly identical $2e^2/h$-quantized conductance peaks, with respect to dot potential.

%
\subsection{Effect of Super Gate Potential}\label{sec:superGate}
%

By sweeping the super gate potential experimentalists are able to adjust the chemical potential throughout the wire. The low energy spectrum as a function of chemical potential (Fig \ref{fig:Fig4}(top)), shows robust zero modes which form in the trivial regime (blue) due to ps-ABSs, well before the TQPT (red). This results in a robust $2e^2/h$-quantized conductance plateau in the ZBCP as a function of chemical potential. As in Ref. \cite{zhang2018quantized} the quantized ZBCP shows oscillatory behavior due to peak splitting in which the ZBCP returns back to $2e^2/h$-quantized value but never exceeds it.
%
\subsection{Effect of Rotation of the Zeeman Field}\label{sec:rotate}
%
\begin{figure}
	\begin{center}
		\includegraphics[width=0.48\textwidth]{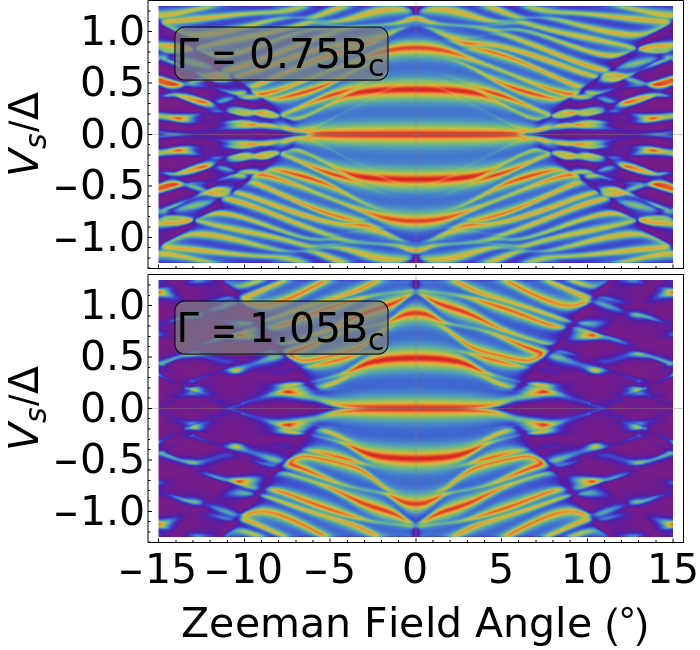}
	\end{center}
	\caption{(Color online) Differential conductance plotted as a function of in plane rotation of the Zeeman field for the ps-ABS (top) and the MZM (bottom) shown in Fig. \ref{fig:wfPlats}. Zero bias peak appears for a small angle in which the Zeeman field is relatively aligned with the wire. As the angle between the wire and the direction of the Zeeman field is increased the ZBCP is destroyed due to splitting.}
	\label{fig:condTheta}
\end{figure}
\begin{figure}
	\begin{center}
		\includegraphics[width=0.48\textwidth]{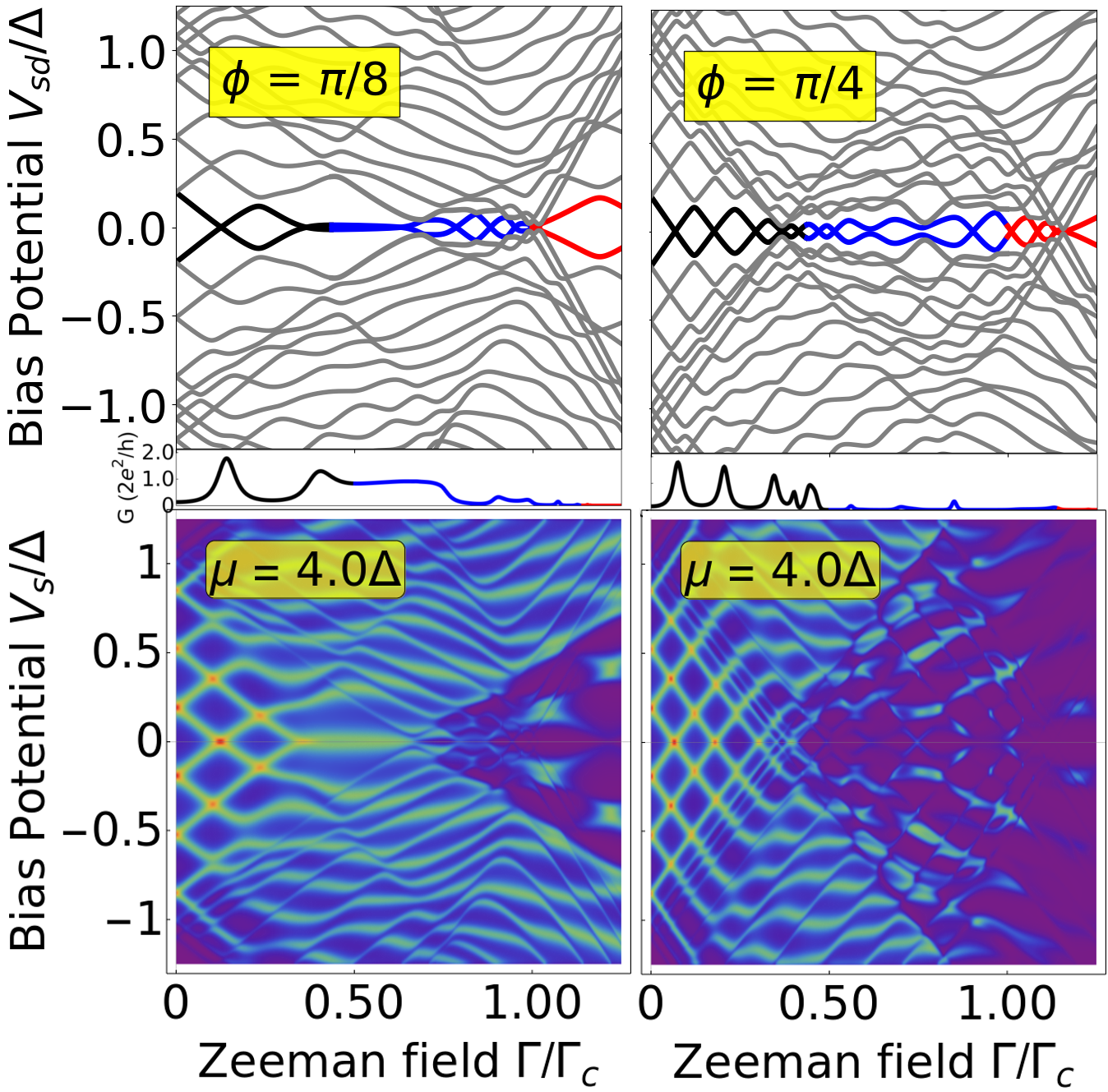}
	\end{center}
	\caption{(Color online)(top)Low-energy spectra as a function of Zeeman field the blue (red) region represents values of the Zeeman field for which the wire supports ps-ABSs (MZMs) at $\phi=0$. In plane rotation of the Zeeman field similarly destroys the ability of both ps-ABSs and MZMs to form stable zero modes. (middle) ZBCP as a function of Zeeman field associated with the plots above. As the magnetic field is increased and the spectrum becomes gapless the $2e^2/h$-quantized conductance plateau is destroyed in both the ps-ABS and MZM regime. (bottom) Differential conductance as a function of Zeeman field associated with the spectrums in the panels directly above.}
	\label{fig:angles}
\end{figure}

SM-SC nanowires require a magnetic field which is aligned with the direction of the wire in order to support topological MZMs. Thus re-orientating the magnetic field should rapidly destroy a ZBCP produced due to a MZM. Fig. \ref{fig:condTheta} shows differential conductance plotted as a function of the angle from the parallel direction in the plane of the wire as in Ref. \cite{zhang2017ballistic} for a trivial ps-ABS (top) and a MZM(bottom). In both plots the ZBCP is only present for a small value of the angle $\phi$ in which the majority of the magnetic field is still aligned with the direction of the wire. For larger rotation angles Fig. \ref{fig:angles} the ZBCP is rapidly destroyed, leaving the system gapless above a critical rotation angle. As a result rotating the magnetic field destroys the quantized conductance plateau due to MZMs as well as due to ps-ABSs Fig.\ref{fig:angles}(middle).

%
\subsection{Scaling of the ZBCP with Temperature}\label{sec:temp}
%

\begin{figure}
	\begin{center}
		\includegraphics[width=0.48\textwidth]{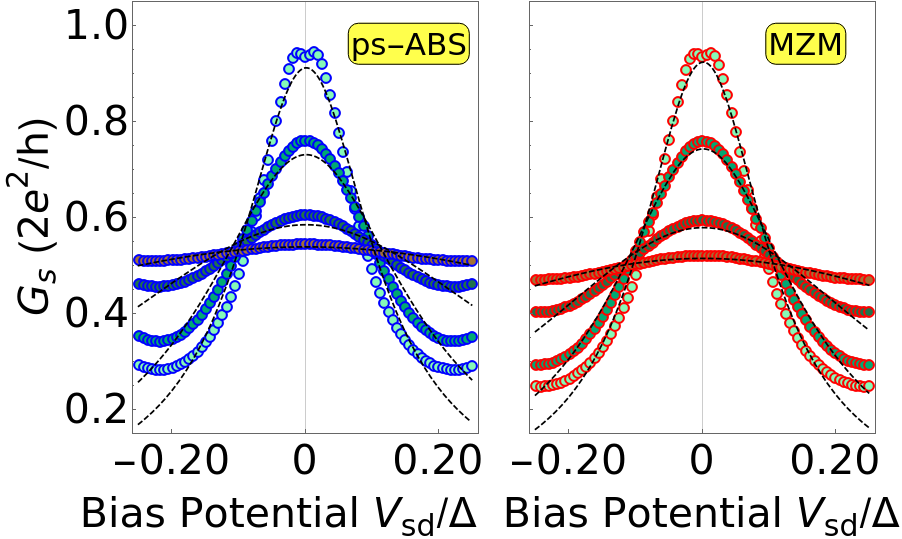}\vspace{3mm}\llap{\parbox[b]{138mm}{\large\textbf{(a)}\\\rule{0ex}{45mm}}}\llap{\parbox[b]{63mm}{\large\textbf{(b)}\\\rule{0ex}{45mm}}}		
		\includegraphics[width=0.23\textwidth]{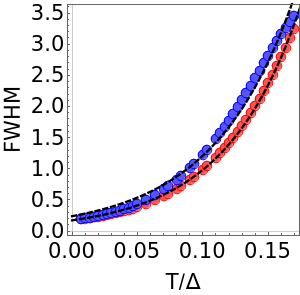}\llap{\parbox[b]{54mm}{\large\textbf{(c)}\\\rule{0ex}{33mm}}}		
		\includegraphics[width=0.21\textwidth]{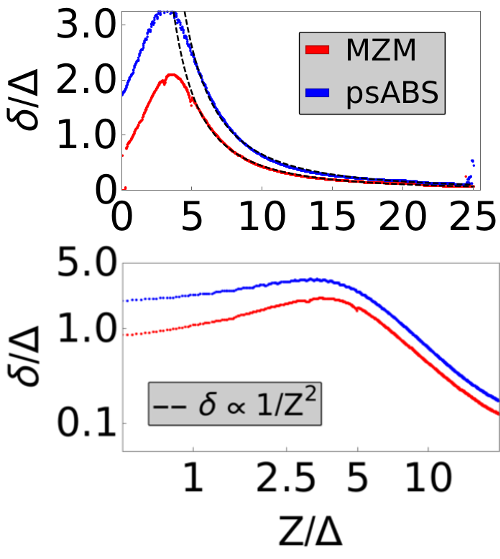}
		\llap{\parbox[b]{51mm}{\large\textbf{(d)}\\\rule{0ex}{28mm}}}
	\end{center}
	\caption{(Color online)(a)-(b) Differential conductance $G_s$ as a function of bias potential $V_{sd}$, for values consistent with ps-ABS and MZM pictured in Fig. \ref{fig:wfPlats}, at various temperatures. (c) FWHM as a function of temperature $T$ for ps-ABS (blue) and MZM (red). Data is fit to an exponential (dotted line)  (d) Linear and log-log plot of intrinsic tunnel broadening $\delta$ as a function of barrier height $Z$ associated with values in Fig. \ref{fig:potential}. In the limit $Z\gg 1$, $\delta\propto 1/Z^2$ for both ps-ABS (blue) and MZM (red).}
	\label{fig:FWHM}
\end{figure}

\begin{figure}
	\begin{center}
		\includegraphics[width=0.48\textwidth]{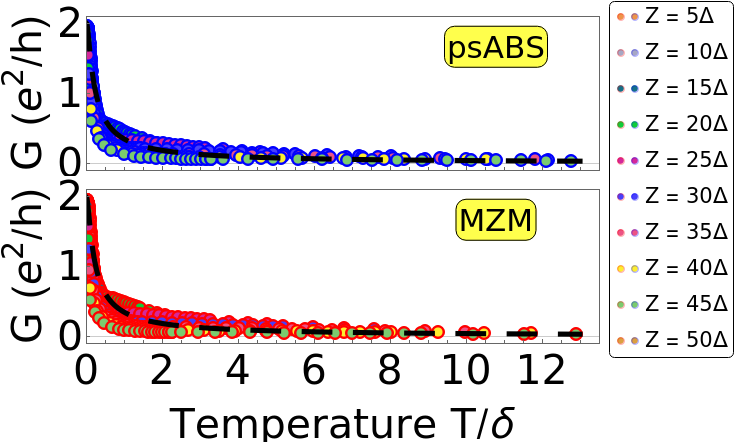}
	\end{center}
	\caption{(Color online) Parameters used correspond to those in Fig. \ref{fig:wfPlats}(b) for ps-ABS and Fig. \ref{fig:wfPlats}(c) MZM. In the weak coupling limit ($Z\gg1$) ZBCP for both the ps-ABS and the MZM scale as a dimensionless ratio of the  temperature and the tunnel coupling.}
	\label{fig:Fig5}
\end{figure}

In Fig. \ref{fig:FWHM} we show the differential conductance as a function of bias potential, taken at various temperatures, for values consistent with the ps-ABS and the MZM shown in Fig. \ref{fig:wfPlats}. The peaks are fit using a Lorentzian function (dotted lines) and the full width at half maximum FWHM (representing the thermal broadening of the ZBCP) is plotted as a function of temperature Fig. \ref{fig:FWHM}(c). The ps-ABS (blue) shows exponential temperature broadening of the ZBCP width similar to that of the MZM (red). The intrinsic tunnel broadening $\delta$ (the FWHM of the ZBCP as $T\rightarrow0K$) for the MZM and the ps-ABS both show  $\delta\propto 1/Z^2$ (Fig. \ref{fig:FWHM}(d)) in the weak coupling limit ($Z\gg 1$). The ZBCPs for both ps-ABSs as well as MZMs are shown in Fig. \ref{fig:Fig5} to scale as the dimensionless ratio 
\begin{align}
	\begin{split}
		G_S(V) &\approx \frac{e^2}{h}\int_{-\infty}^{\infty}dE\frac{2\delta^2}{E^2+\delta^2}\frac{1}{4T\cosh^2(E/(2T))} \\
		&=\frac{2e^2}{h}g(T/\delta)
	\end{split}
\end{align}
of the temperature $T$ and the intrinsic tunnel broadening $\delta$ as in Refs.\cite{setiawan2017electron,nichele2017scaling}.


%
\section{Conclusion}\label{sec:conclusion}
%

By performing  extensive numerical calculations based on tight-binding models of a quantum dot--semiconductor--superconductor structure,
we have shown that partially separated Andreev bound states (ps-ABSs) generate robust zero bias conductance plateaus in end-of-wire charge tunneling experiments, which are indistinguishable from the conductance plateaus generated by non-Abelian MZMs  localized at the  ends of the wire.
In light of these results, we conclude that the recent experimental observations showing $2e^2/h$-quantized conductance plateaus as a function of various control parameters cannot represent definitive evidence for the presence of MZMs. In general, because ps-ABSs are able to produce robust mid-gap modes that stick to zero energy well before the TQPT, resulting in quantized conductance plateaus that are identical to those generated by true MZMs, we conclude that localized tunneling experiments are not sufficient for discriminating between MZMs and ps-ABSs located near the end of the nanowire. We emphasize that a more ``realistic'' modeling (which would face major challenges, considering our limited knowledge of key microscopic parameters that characterize the hybrid systems studied experimentally) will not modify this conclusion. Essentially, when coupling locally to a ps-ABS one couples much stronger to one of the constituent MBSs than to the other, which, basically, remains ``invisible''. Thus, the local coupling to a ps-ABS is effectively equivalent to the local coupling to a MZM.  Discriminating between these types of low energy modes -- hence, demonstrating the realization of non-Abelian, topologically protected Majorana modes localized at the ends of the wire -- requires a non-local probe, such as the two-end charge tunneling measurement. By contrast, our calculations suggest that robustness against control parameter variations  does not prove the presence of true MZMs in any experiment involving only one local probe. Based on the results in this paper, we can conclude that the local charge tunneling measurement, which was, so far, the primary type of probe used in experiments, has exhausted its potential to reveal useful information regarding the distinction of MZMs from low energy ABSs (ps-ABSs in particular) which can appear in SM-SC hybrid structures. The next stage must involve non-local probes, such as, for example, the two-terminal charge tunneling measurement.\cite{moore2018two-terminal,sarma2012splitting}


\section{Acknowledgments}
C.M. and S.T. acknowledge support from ARO Grant
No. W911NF-16-1-0182. T.D.S. was supported by NSF Grant
No. DMR-1414683.


\bibliography{bibFile.bib}

\end{document}